\documentclass[twocolumn]{aastex631}
\usepackage{graphicx, amsmath, amssymb, amsthm}
\usepackage{natbib}
\usepackage{epsfig}
\usepackage{url}
\usepackage{bm}
\usepackage{color}
\begin{document}

\title{Evidence of stellar oscillations in the post-common envelop binary candidate ASASSN-V J205543.90+240033.5}
\author{J..Takata}
\affiliation{Department of Astronomy, School of Physics, Huazhong University of Science and Technology, Wuhan 430074, People's Republic of China}
\author{A.K.H. Kong}
\affiliation{Institute of Astronomy, National Tsing Hua University, Hsinchu 30013, Taiwan}
\author{X.F. Wang}
\affiliation{Department of Astronomy, School of Physics, Huazhong University of Science and Technology,
  Wuhan 430074, People's Republic of China}
\author{F.F. Song}
\affiliation{Yunnan Observatories, Chinese Academy of Sciences, 650011 Kunming, Yunnan Province, People's Republic of China}
\affiliation{Key Laboratory for the Structure and Evolution of Celestial Objects, Chinese Academy of Sciences, 650011 Kunming, People's Republic of China}
\affiliation{University of Chinese Academy of Sciences, 100049 Beijing, People's Republic of China}
\author{J. Mao}
\affiliation{Yunnan Observatories, Chinese Academy of Sciences, 650011 Kunming, Yunnan Province, People's Republic of China}
\affiliation{Center for Astronomical Mega-Science, Chinese Academy of Sciences, 20A Datun Road, Chaoyang District, 100012 Beijing, People's Republic China}
\affiliation{Key Laboratory for the Structure and Evolution of Celestial Objects, Chinese Academy of Sciences, 650011 Kunming, People's Republic of China}
\author{X. Hou}
\affiliation{Yunnan Observatories, Chinese Academy of Sciences, 650011 Kunming, Yunnan Province, People's Republic of China}
\affiliation{Key Laboratory for the Structure and Evolution of Celestial Objects, Chinese Academy of Sciences, 650011 Kunming, People's Republic of China}
\author{C.-P. Hu}
\affiliation{Department of Physics, National Changhua University of Education, Changhua 50007, Taiwan}
\author{L. C.-C. Lin}
\affiliation{Department of Physics, National Cheng Kung University, Tainan 701401, Taiwan}
\author{K.L. Li}
\affiliation{Department of Physics, National Cheng Kung University, Tainan 701401, Taiwan}
\author{C.Y. Hui}
\affiliation{Department of Astronomy and Space Science, Chungnam National University, Daejeon 305-764, Korea}

\email{takata@hust.edu.cn, akong@gapp.nthu.edu.tw}

\begin{abstract}
  ASASSN-V J205543.90+240033.5 (ASJ2055) is a possible  post-common envelope binary system. Its optical photometric data shows an orbital variation
  about $0.52$~days and a fast period modulation  of $P_0\sim 9.77$~minute, whose origin is unknown. In this {\it Letter},
we report an evidence of the stellar oscillation of the companion star as the origin of the fast period modulation. 
  We analyze  the  photometric data taken by  TESS, Liverpool telescope, and  Lulin One-meter Telescope. It is found that the period of the 9.77-minute signal  measured in 2022 August is significantly shorter  than
  that in 2021 July/August, and the magnitude of the change is  of the order of $|\triangle P_0|/P_0\sim 0.0008(4)$. Such a large variation will be
  incompatible with the scenario of the white dwarf spin as the origin of the  9.77-minute  periodic modulation.
   We suggest  that the fast  periodic signal  is  related to  the emission from the irradiated companion star rather
    than that of the white dwarf. Using  existing photometric data covering a wide wavelength range,  we estimate that the hot white dwarf
    in ASJ2055 has a temperature of $T_{eff}\sim 80000$~K and is heating the  oscillating M-type main-sequence star
    with $T_{eff}\sim 3500$~K on its un-irradiated surface. 
    The stellar oscillation of M-type main-sequence star has been predicted in theoretical studies,
    but no observational confirmation has been done.  ASJ2055, therefore,  has a potential to  be  a unique laboratory to investigate the stellar oscillation of a M-type main-sequence star and the heating effect
    on the stellar oscillation.

\end{abstract}

\section{Introduction}
\begin{figure*}
  \epsscale{1}
  \centerline{
    \includegraphics[scale=0.5]{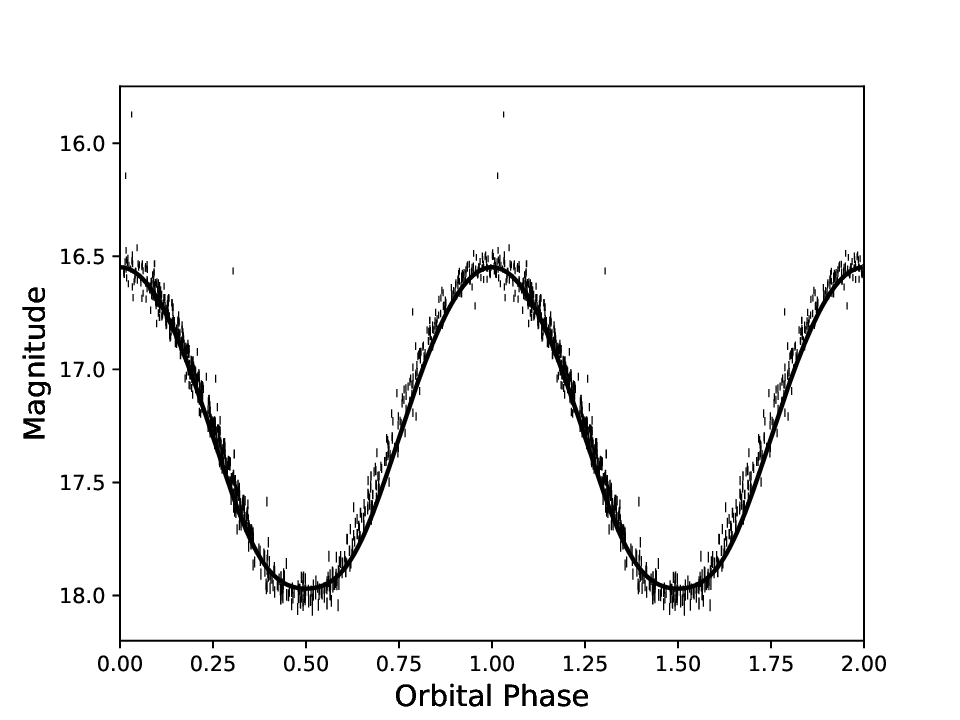}
    \includegraphics[scale=0.5]{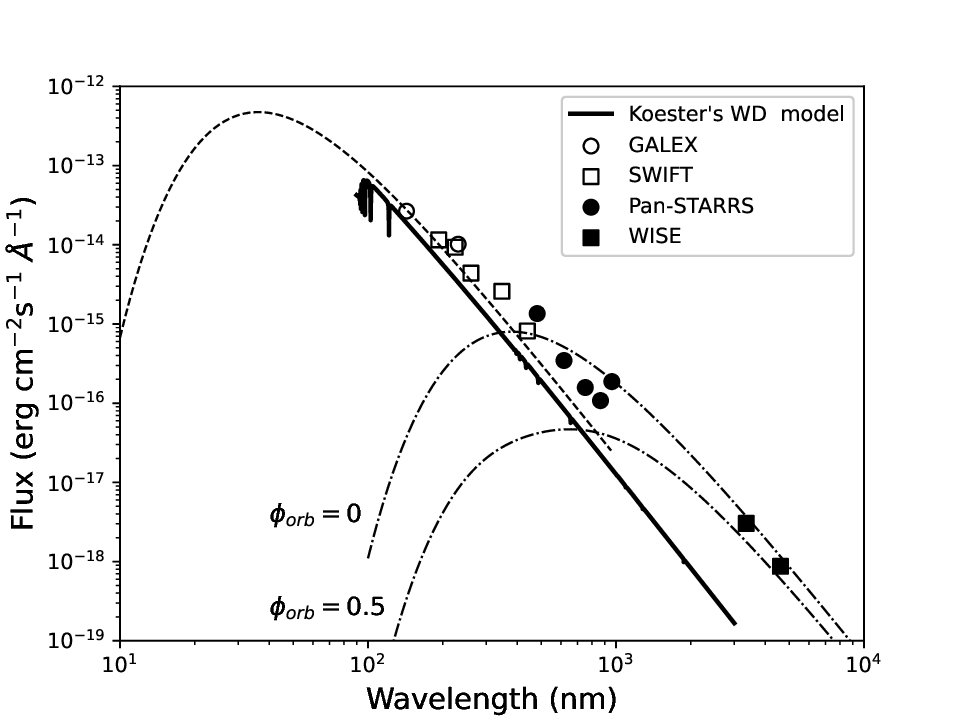}}
  \caption{Left: Orbital modulation of ASJ2055 observed ZTF ($\lambda=750$~nm). The solid line is the model light curve (section~\ref{heatmodel}). Right: Spectrum of ASJ2055. The solid line and dashed line are
    the Koester's WD model \citep{2010MmSAI..81..921K} and the Planck function, respectively, with the effective temperature of
    $T_{eff,wd}=80000$~K, where the radius of WD and
    the distance to the target from  Earth are  assumed to be
    $R_{WD}=1.7\times 10^9$~cm and $d=1.7$~kpc, respectively. The dashed-dotted lines represent
    theoretical spectra of the heated companion star (section~\ref{heatmodel}) at
    the optical peak ($\phi_{orb}=0$) and minimum ($\phi_{orb}=0.5$), respectively.
    For the observed data,  the interstellar dust extinction is calculated using the model of \cite{2007ApJ...663..320F} ($R_V=3.1$)
    implemented in \texttt{extinction} of  Python-code \citep{2016zndo....804967B}; the extinction in the $V$ bands, $A_V=0.5$,
    is estimated from the relation $N_H/A_V\sim 1.8\times 10^{21}~{\rm cm^2mag^{-1}}$ with $N_H\sim 9\times 10^{20} ~{\rm cm^2}$ inferred
    from the sky position of ASJ2055 (\url{https://heasarc.gsfc.nasa.gov/cgi-bin/Tools/w3nh/w3nh.pl}).  Error bar of each
      observational point is smaller than the size of each symbol.  }
  \label{asj}
\end{figure*}

ASASSN-V J205543.90+240033.5 (hereafter ASJ2055)  is a binary system, which is composed of a hot white dwarf (WD)
and a cool main-sequence star that is  detached from the Roche-lobe. 
The information of the binary nature of ASJ2055 is  reported by \cite{2021arXiv210809060K} and \cite{2021arXiv210903979K}, who find
two periodic modulations  with $\sim 0.5$~day and $\sim 9.77$~minute in the optical data taken by the Zwicky Transient
Facility  \citep[hereafter ZTF, ][]{2019PASP..131a8003M}. The former is thought to be the orbital period ($P_{orb}$),
while the origin of the latter ($P_0$) has not been understood.  An interesting property of ASJ2005 is that  the orbital light curve in  the optical bands  shows a single broad peak  with a large amplitude of $\triangle m\sim 1.5$~magnitude (Figure~\ref{asj}).  This orbital modulation 
is interpreted as a result of the irradiation on day-side
of the companion star by the WD \citep{2021arXiv210809060K, 2021RNAAS...5..242W}, and a rate of energy deposited on the companion star will
be of the order of $\sim 10^{32}~{\rm erg~s^{-1}}$.

If the periodic signal with the 9.77-minute signal  represents a spin period of the WD, ASJ2055  may be a binary system similar to AR~Scorpii \citep{2016Natur.537..374M,2022MNRAS.516.5052P}, as suggested by \cite{2021arXiv210809060K}.
AR~Scorpii comprises a  WD and a low-mass (M-type)  companion star, and its orbital period is $P_{orb}\sim 3.56$~hours.
It also shows a large orbital variation ($\triangle m\sim 2$~magnitude) in the optical light curve 
and contains a rapidly spinning WD with a spin-period of $P_s\sim 118$~seconds. Signature of the non-thermal emission in broad energy bands
from radio to X-ray \citep{2016Natur.537..374M, 2017NatAs...1E..29B, 2018ApJ...853..106T,2022MNRAS.510.2998D} of AR~Scorpii suggests a particle acceleration process in the WD binary system.

Although the optical properties of two binary systems are similar to each other,  the heating process of the companion star in
ASJ2055 may be  different from that in  AR~Scorpii.
For AR~Scorpii,  the temperature of the WD's surface
  is $\sim 11,500$~K, indicating that WD luminosity level, $L_{WD}\sim 10^{31}~{\rm erg~s^{-1}}$,  is lower  than  the luminosity $\sim 10^{32}~{\rm erg~s^{-1}}$ of the companion star \citep{2016Natur.537..374M, 2021ApJ...908..195G}.  It is, therefore,  suggested  that  AR Scorpii contains a fast spinning magnetized WD with a surface magnetic field of $B_s\sim 10^{7-8}$~Gauss, and  the WD's magnetic field  or rotation will  be the energy source of the irradiation and non-thermal activities \citep{2016Natur.537..374M, 2017NatAs...1E..29B,2016ApJ...831L..10G, 2017ApJ...851..143T,2018MNRAS.476L..10B,2020arXiv200411474L}. For ASJ2055, \cite{2021RNAAS...5..242W} measures the spectrum in $\sim 300-1000$~nm bands, and find that the flux is rising steeply toward the UV bands (see Figure~\ref{asj}).
With the property of the spectrum, they  suggest  that ASJ2055 is a post-common envelop binary (hereafter PCEB)  and contains a hot WD that heats up the
companion star, which is probably a M-type star. Hence, the origin of the 9.77-minute periodic modulation has not been well understood.

In this {\it Letter}, we carry out a more detailed photometric study   to probe    the origin of
the  9.77-minute periodic signal.
The structure of this paper is as follows. We describe the data reduction  in section~2.
We present the results of the timing analysis of the photometric data in sections~\ref{stability} and~\ref{orbit},
and modeling for the orbital modulation of the  light curve in section~\ref{heatmodel}.
In section~4, we suggest the oscillation of the companion  star is the origin of 9.77-minute periodic signal, and AJS2055
is a new type PCEB, in which 
 a hot WD heats up the oscillating low-mass main-sequence star.
\section{Data reduction}

We analyze photometric data taken by ZTF, Transiting Exoplanet Survey Satellite~\citep[hereafter TESS,][]{2014SPIE.9143E..20R}, Lulin One-meter Telescope (hereafter LOT) in Taiwan, two-meter Liverpool telescope (hereafter LT) in Spain and the Neil Gehrels SWIFT Observatory~\citep[hereafter SWIFT,][]{2005SSRv..120..165B}. For ZTF data, we  download
the light curves from the Infrared Science Archive\footnote{\url{https://irsa.ipac.caltech.edu}},
and use the data (DR8 object) in $r$-band to determine the orbital period (Figure~\ref{asj}).  TESS observed ASJ2055
in 120-second cadence mode \citep{TESSALL} in  2021 July/August and 2022 August.
We download the light curves from Muikulski Archive for Space Telescopes (MAST) Portal\footnote{\url{https://mast.stsci.edu/portal/Mashup/Clients/Mast/Portal.html}} and use Pre-search Data Conditioning Simpler Aperture Photometry (PDCSAP) flux to analyze the light curve. The top two panels in Figure~\ref{tess} present the temporal evolution of PDCSAP flux, and the light curves  modulate with the orbital period of $P_{orb}\sim 0.52$~day. We find in the figure that 
  the  light curve generated by the TESS SAP pipeline becomes negative values around the optical minimum.
  We investigate the origin of the negative flux by performing a custom
  aperture photometry\footnote{\url{https://heasarc.gsfc.nasa.gov/docs/tess/Target-Pixel-File-Tutorial.html}} with  the target pixel files (hereafter TPFs).  We obtain 
  the TPFs from TESS Science Processing Operations Center\footnote{\url{https://heasarc.gsfc.nasa.gov/docs/tess/}} and analyze them   using {\it lightkurve} tool in Python\footnote{\url{https://docs.lightkurve.org/tutorials/2-creating-light-curves/2-1-cutting-out-tpfs.html}}.
  From the TPFs, we find that ASJ2055 is a faint source and the differential light curve (i.e. the  light curve of the source pixels 
  minus the background pixels) can have  negative values in the light curve. We also confirm that
  the shape of the differential light curve is not significantly affected by the
  choice of the background region. 
  Therefore, we use the pipeline-generated light curve in this study.

We carried out   LOT and LT observations for several nights in 2021 and 2022; exposure
length for each night is from several ten minutes to several hours.
Table~1 summarizes the  information of the LOT and LT observations. With the extracted photometric light curve,  we create a  Lomb-Scargle periodogram \citep[hereafter LS,][]{1976Ap&SS..39..447L} to search for a periodic modulation in the light curves, and we estimate false of alarm probability (FAP) using  the methods of \cite{2008MNRAS.385.1279B}.

SWIFT observed ASJ2055 with a total exposure of $\sim 22$~ks. We extract the clean event files for
UVOT and   XRT  data using \verb|HEASoft ver.6-29|.  For UVOT data, we extract the light curve and magnitude  with command \verb|uvotevtlc| and \verb|uvotsource| in \verb|HEASoft|, respectively.
For XRT data, we extract image with \verb|Xselect| and confirm no significant emission
at the source position, which is consistent with the result reported by \cite{2021ATel14932....1G}.
We estimate $F_X\sim 5\times 10^{-14}~{\rm erg~cm^{-2}s^{-1}}$ as the 3-$\sigma$ upper limit
using the command \verb|uplimit| in \verb|XIMAGE| package (version 4.5.1). 

\section{Results}
\label{result}
\subsection{Stability of the  9.77-minute signal}
\label{stability}
\begin{figure}
  \epsscale{1}
  \centerline{
    \includegraphics[scale=0.6]{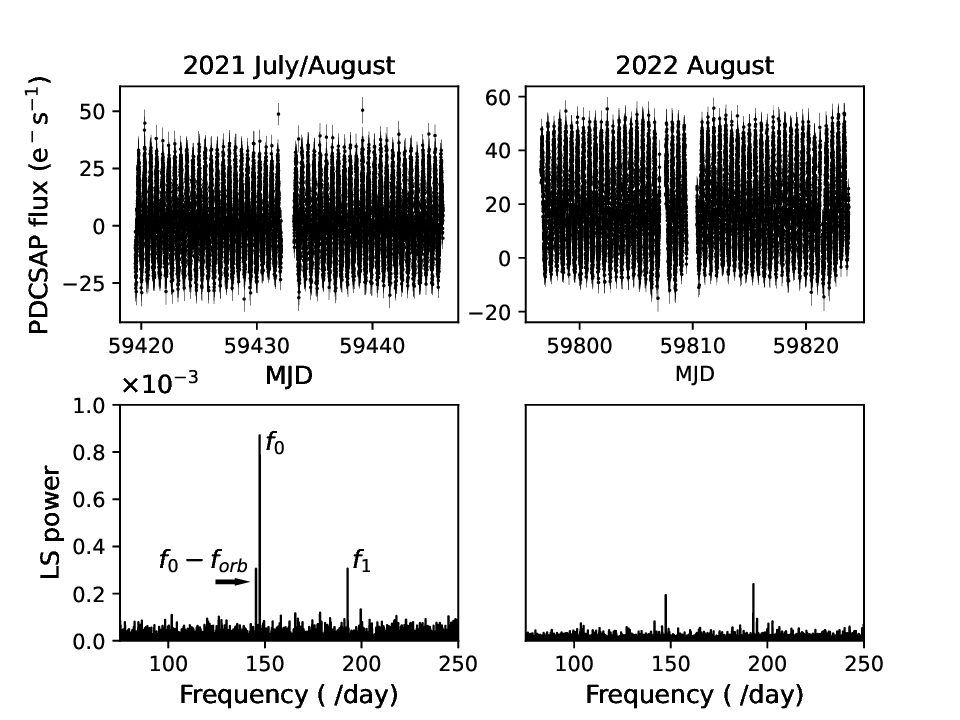}}
  \caption{Top panels: TESS-PDCSAP flux for AJ2055 taken in 2021 July/August (top left) and 2022 August (top right) downloaded from MAST. Bottom panels:
    The LS-periodogram  of  2021 (left panel) and  2022 (right panel) TESS data after subtracting the mean value of the light curve.
    The frequency range of $75-250~{\rm day^{-1}}$ is
      displayed to present the signals of $f_0\sim 147~{\rm day^{-1}}$ and $f_1\sim 192~{\rm day^{-1}}$. The frequency $f_0-f_{orb}$
      corresponds to  the beat signal.}
  \label{tess}
\end{figure}

\begin{figure}
  \epsscale{1}
  \centerline{
    \includegraphics[scale=0.65]{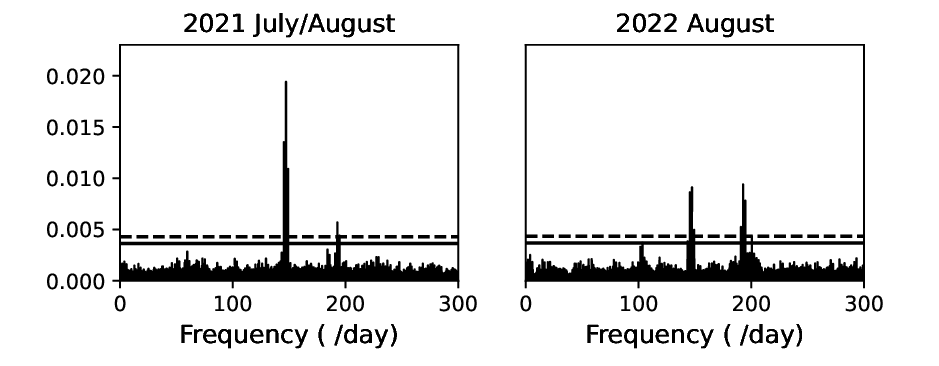}}
  \caption{LS-periodogram for 2021 July/August (right) and 2022 August TESS data after subtracting the orbital modulation and the data     taken at the orbital phase
    of $\phi_{orb}=0-0.2/0.8-1$, where
    the periodic signal can be confirmed in the TESS data.
    The orbital modulation is removed from the light curve.
    The solid lines and dashed lines correspond to  FAP=0.1 and 0.01 estimated
    by the methods of \cite{2008MNRAS.385.1279B}, respectively.}
  \label{tess1}
\end{figure}

\begin{figure}
  \epsscale{1}
  \centerline{
    \includegraphics[scale=0.65]{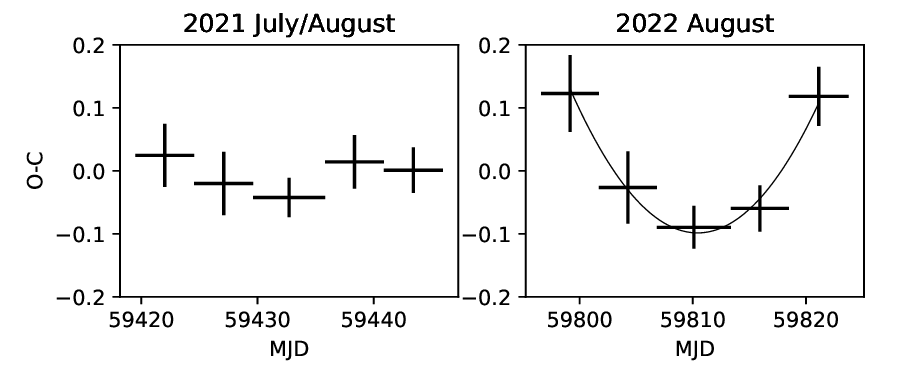}}
  \caption{The O$-$C curves for  the  arrival phase  of the $9.77$-minute periodic signal
      using the TESS data. The light curve  is folded with the
      frequency of $f_0=147.28~{\rm day^{-1}}$ for 2021 July/August data (left panel) and $f_0=147.40~{\rm day^{-1}}$ for 2022 August
      data (right panel). A Gaussian function is applied to fit the  pulse profile.
      In the right panel, the solid curve is the fitting
      of the O$-$C curve and indicates the first time derivative of the period of $\dot{P}_0=1.2(5)\times 10^{-7}$.}
  \label{phase}
\end{figure}

Figure~\ref{tess} shows the light curves of  TESS-PDCAP flux for ASJ2055 taken in 2021 July/August (top left) and 2022 August (top right), and clearly indicates a modulation with the orbital period of $P_{orb}\sim 0.523$~day,  whose frequency signal ($f_{orb}\sim 1.91~{\rm day^{-1}}$)  dominates the LS-periodogram of the TESS light curve.
  Figure~\ref{tess} displays the LS-periodogram in the
  frequency range of 75~$\rm{day^{-1}}$-250~$\rm{day^{-1}}$, and shows 
  a  short periodic signal of $P_0\sim 9.77$~minute ($f_0\sim 147~{\rm day}^{-1}$).  In addition to
  the signal $f_0$, we can also confirm the existence of the  beat signal
  at $f_0-f_{orb}\sim 145~{\rm day^{-1}}$ in 2021 data and a periodic signal at $f_1\sim 192~{\rm day^{-1}}$. The beat signal will be 
  related to the fact that in the current TESS data, the short periodic signal can be confirmed during the orbital phases $\phi_{orb}=0-0.2/0.8-1$, where $\phi_{orb}=0$ corresponds to the optical peak in Figure~\ref{asj} (section~\ref{orbit}).

    To remove the effect of the orbital modulation from the light curve, we fit the light curve of Figure~\ref{tess} with a functional  form
    of $\mathcal{F}=c_0\sin[2\pi (ft+c_1)]+c_2t+c_3$, where the first term is the periodic modulation, and the second
    and third terms correspond  to the linear trend and the  base of the light curve, respectively.
    In addition,  $f$ is the peak frequency in LS-periodogram, and it is the orbital frequency
at  the first iteration in the pre-whitening process. Then, we  subtract
its contribution from  the light curve and create  a new  LS-periodogram to find a new peak frequency.
We iterate this process until no significant signal with $f<10~{\rm day^{-1}}$ appears in the LS-periodogram.
Figure~\ref{tess1} shows the LS-periodogram after removing the orbital modulation with TESS data taken at
the orbital phase $\phi_{orb}=0-0.2/0.8-1$. We can see that significance of signals increases, and  we find a marginal third
signal at  $f_2\sim 103~{\rm day^{-1}}$ in 2022 August data (right panel in Figure~\ref{tess1}).

We find that  two detected frequencies $f_0=147.28(4)~{\rm day^{-1}}$ in 2021 data
and $147.40(4)~{\rm day^{-1}}$ in 2022 data are significantly different from  each other; the error is estimated from the Fourier width (the inverse of  total exposure).  Such a large change of the periodic
signal $|\triangle  P_0|/P_0\sim  0.0008(4)$ in a scale of year will not be realized by  the change of the spin period of the WD. This suggests that the WD's
spin as the origin of  $P_0\sim 9.77$~minutes periodic signal is unlikely.

We create an observed-minus-computed (hereafter O$-$C) curves of the arrival phase of the pulsed peak with the TESS data (Figure~\ref{phase}), and investigate the day/month timescale stability of the signal $f_0$. With the TESS light curve after removing the orbital modulation,
we determine the averaged frequency and arrival phase of the pulse of the  2021 or 2022 TESS data set. Then, we divide 2021 or 2022 data  into five segments and fold  each epoch  with the averaged frequency. We determine the arrival phase of the pulsed peak and calculate the difference
from the averaged one. Figure~\ref{phase} shows the O$-$C curves for 2021 data (left panel) and 2022 data (right panel). We can see  that the 2021 data does not show a significant temporal variation of the arrival phase of the pulsed peak. In 2022 August data (right panel), on the other hand,
we may see a temporal variation of the 9.77-minute periodic modulation with  the 
first time derivative of the
period of $\dot{P}_0=1.2(5)\times 10^{-7}$, which can be also confirmed
in 2-dimensional LS-periodogram.  If it would correspond  to the spin down of the WD,
the spin down energy were  $I_{WD}(2\pi)^2P^{-3}_0\dot{P}_0\sim 10^{37}~{\rm erg~s^{-1}}$ with $I_{WD}\sim 10^{51}{\rm g~cm^{2}}$ being  the moment of inertia of the WD. Such a unrealistically large spin down rate also indicates that the WD's spin is unrelated to the 9.77-minute periodic signal.   

In LS-periodogram, the second periodic signal is confirmed  at $f_1\sim 192.77(4)~{\rm day^{-1}}$ ($P_1\sim 7.47$~minute)
  for 2021 data and at $f_1\sim 192.75(4)~{\rm day^{-1}}$ for 2022 data. This second periodic signal would
  not be explained by a simple harmonic of $f_0\sim 147~{\rm day^{-1}}$ signal or the combination of $f_0$ and the orbital frequency $f_{orb}$. Moreover,
  we can see in  2022 data that the power of $f_1$-signal is comparable to that of  $f_{0}$-signal. We expect that $f_1$-signal has  a different mode of the stellar oscillation of the low-mass
  companion star (section~\ref{discuss}).

\begin{table}
  \caption{Journal of LOT and LT observations. The zero orbital phase  is defined at MJD~58957}
  \begin{tabular}{cccc}
    \hline
    & Date & Orbital phase  & Filter \\
    \hline\hline
    LOT  & 2021/10/04 & 0.14-0.36 & $r$\\
    & 2021/10/05 & 0.05-0.24 & $r$\\
    & 2022/08/14 & 0.34-0.51 & No\\
    & 2022/08/15 & 0.30-0.43 & No  \\
    & 2022/10/01 & 0.74-0.87 & No  \\
    & 2022/10/02 & 0.43-0.51 & No  \\
    &       & 0.75-0.81 & No  \\
    & 2022/10/13 & 0.59-0.67 & No  \\
    & 2022/11/14 & 0.60-0.75 & No  \\
    & 2022/11/15 & 0.52-0.66 & No  \\
    & 2022/11/16 & 0.42-0.56 & No  \\
    & 2022/11/17 & 0.33-0.47 & No  \\
    & 2022/11/18 & 0.24-0.38 & No  \\
    \hline
    LT    & 2022/08/03 & 0.58-0.63 & $r$\\
    & 2022/08/08 & 0.31-0.36 & $r$\\
    & 2022/08/11 & 0.95-1.0 & $r$\\
    & 2022/08/18 & 0.43-0.49 & $r$\\
    & 2022/08/20 & 0.47-0.51 & $r$\\
    & 2022/09/06 & 0.56-0.59 & $r$\\
    \hline
  \end{tabular}
\end{table}

\subsection{Orbital variation of the 9.77-minute signal}
\label{orbit}
\begin{figure*}
  \epsscale{1}
  \centerline{
    \includegraphics[scale=0.65]{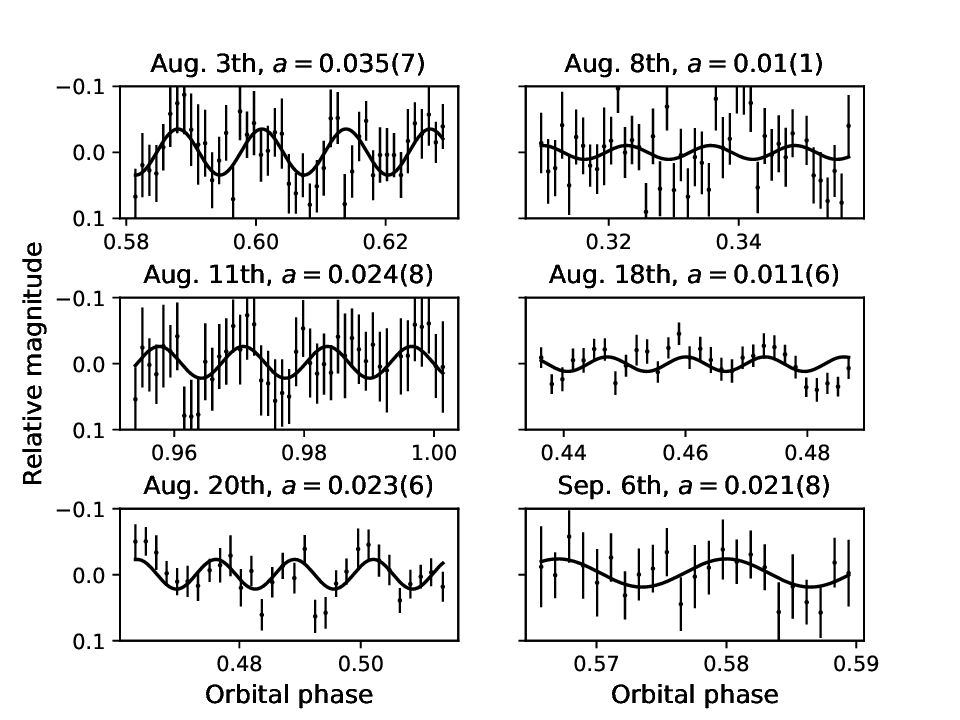}
    \includegraphics[scale=0.65]{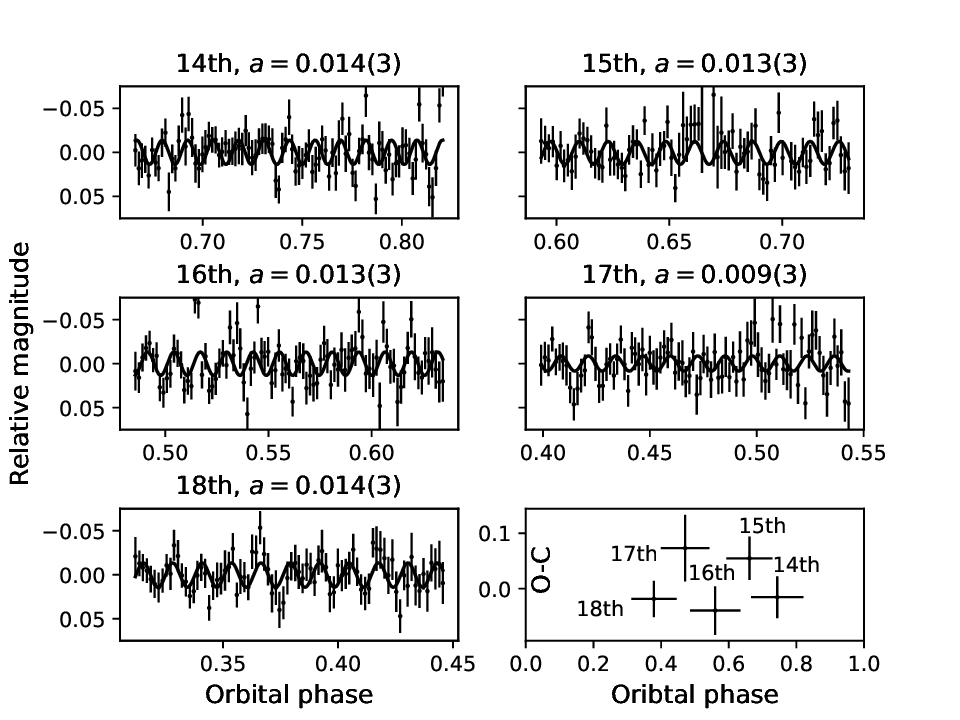}}
  \caption{LT (left) and LOT (right) light curves folded with the orbital period  for the data taken in 2022 August and  November,
    respectively.  The orbital modulation is removed from
    the data.  The solid line shows  a sinusoidal function with a modulation frequency of
      $f_0=147.40~{\rm day^{-1}}$ for LT data and  $f_0=149.35~{\rm day^{-1}}$ for LOT data. For LOT data, the O$-$C curve of the
      phase of the pulsed peak is presented. The calculated amplitude ($a$) of the modulation
      for each data set is presented at the top of each panel. }
  \label{lulin}
\end{figure*}

  The existence of the beat signal, $f_0-f_{orb}$, indicates an orbital variation of the amplitude  of
  the 9.77-minute signal.
  In TESS observation, we only confirm the significant periodic signal  near the optical peak (the orbital phase $\phi_{orb}\sim 0-0.2/0.8-1$ in the left panel of Figure~\ref{asj}). 
  We carried out LOT and LT observations in 2021 and 2022 to investigate the orbital variation of the periodic signal.
  The left panel of Figure~\ref{lulin} shows the light curves folded with the orbital period for 2022 August  LT data and the right panel shows the light curve of  November
  LOT data. Both data sets cover  the optical minimum of the orbital light curve, and  clearly indicate  a modulation with a 9.77~minute. We confirm that the 9.77-minute periodic modulation exists for the entire orbital phase.

  We investigate the orbital variation of the 9.77-minute  periodic signal with 2022 August LT and 2022 November LOT observations. For LT data, since we  cannot obtain a significant periodic signal in LS~periodogram, we apply $f_0=147.40(4)~{\rm day^{-1}}$ obtained with the 2022 August TESS data. For  November LOT data, we obtain $f_0\sim 149.6(2)~{\rm day^{-1}}$ 
  as the best frequency in LS-periodogram. We use a sinusoidal function to  determine its amplitude
  and the peak phase  of the modulation.  In Figure~\ref{lulin}, we present the O$-$C values of the peak phase for LOT data and the amplitude ($a$) for the data of each night.  We can see that within the errors, the O$-$C values and the amplitudes are  consistent with constants throughout the orbital phase. This may suggest that the non-detection of the periodic signal in the TESS data during the orbital phase  $\phi_{orb}\sim 0.2-0.8$  is due to a low signal-to-noise ratio. With the large size of the error bars of
  the current LT/LOT data, however, a deeper observation is also  desired to obtain a more solid conclusion
  for the 
  orbital variation of the 9.77-minute periodic signal.

\subsection{Modeling for orbital modulation}
\label{heatmodel}
\begin{table}
  \caption{Fitting parameters for ASJ2055. $f_{roche}$ represents the Roche-lobe filling factor of the companion star, and $T_{eff,0}$ is the effective temperature of the unheated surface of the companion star. $\theta_0$ is the observer viewing angle measured from the direction perpendicular to the orbital plane.}
  \begin{tabular}{cc}
  \hline
  $d$ & 1.7~kpc \\
$M_{WD}$ & $0.6M_{\odot}$ \\
  $R_{WD}$ & $0.024R_\odot$ \\
  $T_{eff, WD}$ & 80000~K \\
  $M_c$ & $0.3M_{\odot}$ \\
  $f_{roche}$ & 0.5 \\
  $T_{eff,0}$ & 3500~K \\
  $\theta_o$ & $37^{\circ}$ \\
  \hline
  \end{tabular}
\end{table}
ASJ2055 shows a  large amplitude ($\triangle m\sim 1.5$ magnitude) of the orbital modulation in the optical bands. This implies that the companion star, which is probably M-type main-sequence  star, is heated up by the irradiation of the WD.  \cite{2021RNAAS...5..242W}  measure the spectrum of ASJ2055 in 300-1000~nm bands and find that the spectrum rises steeply toward the shorter wavelength.  They suggest that ASJ2055 is a post-common envelope binary (hereafter PCEB) and a hot WD irradiates the day-side of the secondary companion star.  We produce a broadband  spectrum in 100-10000~nm bands using new SWIFT UVOT data and archival GALEX \citep{2010AJ....140.1868W},  Pan-STARRS \citep{2017yCat.2349....0C} and WISE data \citep{2012yCat.2311....0C} \footnote{\url{https://vizier.cds.unistra.fr/viz-bin/VizieR}}, and we confirm  that the spectrum is continuously rising toward shorter wavelength bands below 300~nm~(Figure~\ref{asj}). By assuming that  the fluxes in $\lambda  <300$~nm bands is dominated by the emission from the WD,
we fit the spectrum with a Planck function (dashed line in Figure~\ref{asj}) with an  effective temperature of $T_{eff,WD}=80000$~K and $R_{WD}=1.7\times 10^{9}$~cm,  where we apply the distance to the source of $d=1.7$~kpc measured by  GAIA \citep{2021A&A...649A...1G,2022yCat.1360....0G}. We also generate Koester's WD atmosphere model \citep{2010MmSAI..81..921K} with the gravity acceleration of  ${\rm log}g=7.25$\footnote{\url{http://svo2.cab.inta-csic.es/theory/newov2/}}.

We carry out a modeling for the orbital light curve of the companion star heated by the WD. Because of the large amplitude of the orbital modulation,  the  rotational motion of  the companion star is
likely synchronized with the orbital modulation. Only one surface will be
heated up by the irradiation from the WD. We  take into account  the gravity darkening  as $T_{eff,c}=T_{eff,0}(g/g_{max})^{\beta}$, where $g_{max}$ is the maximum gravity and 
$\beta\sim 0.08-0.25$  \citep{1967ZA.....65...89L,2012A&A...547A..32E}. In this study, we apply $\beta=0.15$ in our calculation.
By assuming that all irradiated flux from the WD is used for heating of the companion star, we may calculate the temperature ($T_{new}$)  for a surface segment of the companion star as
\begin{equation}
  \sigma_{sb}T_{new}^4\delta A=\sigma_{sb}T_{eff,c}^4\delta A+\frac{L_{WD}}{4\pi \ell^2}\delta A_{irr},
\end{equation}
where $L_{WD}$ is the WD's luminosity, $\sigma_{sb}$ is  the Stephen-Boltzmann constant and $\delta A$ is the area of the surface segment, $\ell$ is the
distance between the WD and the surface segment, and $\delta A_{irr}$ is the area of the surface segment measured from the WD.
For each segment, we assume a Planck function for  the emission, and apply a simple limb-darkening effect with
$I(\cos\theta)\propto 1-0.5(1-\cos\theta)$ \citep{1993AJ....106.2096V}, where $\theta$ is the viewing angle of the observer measured from the direction perpendicular to the surface. Table~2 summarizes the parameters of the fitting.
Since no sign of the accretion process is observed, the companion star will be detached from the Roche-lobe, namely,  the Roche-lobe filling factor is less than unity $f_{roche}<1$. To explain the flux level observed by WISE, we choose
the base surface temperature of $T_{eff,0}= 3500$~K. In Figure~\ref{asj}, we compare the results of the model with the observations. We find that the heating due to  the emission from the  WD surface
can explain the observed light curve and  spectrum, simultaneously.

We note that  the effective temperature $(T_{eff,0}$) of the companion
star and the observer viewing angle $\theta_0$ are degenerated in the fitting process for the orbital modulation, and  a reasonable fit can be obtained in
a parameter range of $T_{eff,0}\sim 3300-4500$ and $\theta_o<60^{\circ}$, for which a smaller effective temperature
corresponds to a smaller viewing angle. To explain the flux level of the WISE data,
on the other hand, a smaller temperature is preferred. Hence, a  more detailed spectroscopic information such as a radial velocity curve and the spectral information at night-side of the companion star will be desired to
reveal  the property of the companion star.

\section{Discussion and summary}
\label{discuss}
\subsection{Origin of the 9.77-minutes signal}
  It was suggested that the pulsation  with a frequency
  $f_0\sim 147~{\rm day^{-1}}$ ($P_0\sim 9.77$~minutes) is related to the spin-period of the WD of ASJ2055 \citep{2021arXiv210903979K}. As we discussed in section~\ref{stability}, however,   
  the photometric data indicates a significant change in the period
  $|\triangle P_0|/P_0\sim 8\times 10^{-4}$ between  2021 and 2022, and 
  such a large change  will not be explained by the temporal variation  of the WD's spin.
  With the current results, therefore, we conclude that the  9.77-minute signal is irrelevant  to the WD's spin. It is unlikely that the periodic modulation is related to the stellar oscillation of the  WD. This is because if the periodic
  emission is from the surface of the WD,
 we expect the detection of  the periodic signal in TESS data taken at around the optical minimum, where the contamination of the radiation  from the companion star is minimum. 

 One  possible origin of the 9.77-minute signal is the solar-type oscillation of
 the main-sequence star (\cite{2019LRSP...16....4G} for a recent review).  In addition to the significant
 periodic signals of  $f_0\sim 147.4~{\rm day^{-1}}$ and  $f_1\sim 192.8~{\rm day^{-1}}$, we  also find a marginal third signal at
 $f_2\sim 103~{\rm day^{-1}}$ in LS-periodogram (bottom right panel of Figure~\ref{tess1} ).
  Because these three frequencies are almost equally separated with  $\triangle f =f_0-f_2\sim f_1-f_0\sim  45~{\rm day^{-1}}$,  we may anticipate the main frequency $f_0$ and the separation $\triangle f$ as the so-called  frequency of the maximum power and large frequency separation, respectively, of the solar-type oscillation.  It has been known that the frequency of the maximum power  and
the large separation of the solar-type oscillations are well scaled as \citep{2011A&A...530A.142B, 2020FrASS...7....3H}
\begin{equation}
  f_{max}=f_{max,\odot}\left(\frac{g}{g_\odot}\right)\left(\frac{T_{eff,\odot}}{T_{eff}}\right)^{1/2},
  \label{fmax}   
\end{equation}
and 
\begin{equation}
  \triangle f=\triangle f_{\odot}
  \left(\frac{M}{M_{\odot}}\right)^{1/2}\left(\frac{R}{R_\odot}\right)^{-3/2},
  \label{large}
\end{equation}
respectively,  where  $f_{\odot}=3100~\mu{\rm Hz}$,  $\triangle f_{\odot}=135.1~\mu{\rm Hz}$, $T_{eff,\odot}=5777$~K and ${\rm log}g_{\odot}=4.4377$ are the solar values.

With $M_c\sim 0.3M_{\odot}$, $R_c\sim0.3R_{\odot}$ and $T_{eff}\sim 3500$~K shown in  Table~2, the equations~(\ref{fmax}) and~(\ref{large}) expect $\triangle f\sim 450~\mu H$ and $f_{max}\sim 4000\mu Hz$, respectively. We find that the observed  separation $\triangle f\sim  45~{\rm day^{-1}}\sim 520\mu{\rm Hz}$ of ASJ2055 may be consistent with the scaling law, but  the observed peak  frequency   $f_0\sim 147.4~{\rm day^{-1}}\sim 1700~\mu{\rm Hz}$
is significantly smaller. The day-side of the companion star is heated up to a temperature of $T_{eff}\sim 8000$~K,
with which the scaling law expects $f_{max}\sim 2640\mu {\rm Hz}$.  Hence, the heating of  the companion star  surface may affect to the frequency  of the stellar oscillation.

 Asteroseismology is a powerful tool to investigate the stellar  interior structure  and can provide fundamental stellar parameters with high
precision [e.g. using scaling laws of equations~(\ref{fmax}) and~(\ref{large})].  The stellar oscillation of M-type star
has been theoretically predicted \citep{2014MNRAS.438.2371R,2021MNRAS.507.5747B},
and the efforts have been made to detect the stellar oscillation of the M-type star \citep{2011AcA....61...37B,2011AcA....61..325B,2016MNRAS.457.1851R,2017MNRAS.469.4268B}. However no solid confirmation of the existence of the stellar oscillation of M-type main-sequence star has been done;
one possible signal from  pre M-type main-sequence star has been reported by \cite{2021A&A...654A..36S}. Hence the method of asteroseismology has not been applied to M-type main-sequence star. If the 9.77-minute periodic signal discussed in this study originates from the stellar oscillation, ASJ2055  provides the evidence of the stellar oscillation of the M-type star.
Further investigation will be important  to obtain a solid conclusion about  the origin of the short periodic signal and to identify the stellar
type of the companion star.

\subsection{Other PCEB}

With the launch of TESS, the population of the binary system containing a oscillating star has been increasing \citep{2022ApJS..259...50S,2022ApJS..263...34C}. However, the population of the WD binary system that contains
an oscillating secondary is not many, and the companion star is usually subdward B star \citep{2011MNRAS.412..371R,2018MNRAS.474.4709K, 2022ApJ...928L..14J}. If the 9.77-minute signal is originated from
the stellar oscillation, ASJ2055 will be a new type of PCEB, in which a  hot WD heats up
a oscillating M-dwarf star. ASJ2055 is a relatively young binary system after
the common-envelope phase  and  the cooling timescale of the WD's surface is of the order of
several million years.  Such a PCEB with a hot WD  will be a unique laboratory  to study the pulsating M-type star and the effect of the heating.  We, therefore, carry out  a search for other  candidates of PCEB that contain a oscillating  secondary.
\cite{2010MNRAS.402..620R} provides  $\sim 3300$  of PCEB candidates from SDSS data. From the catalog, we select 25 binary systems, in which temperature of the WD is
greater than $T_{eff}>70000$~K, and we check a short periodic signal in the
archival photometric data downloaded from MAST portal. However, we obtained null results.
This means that ASJ2005 is the rare PCEB system. 

We mention  SDSS J082145.27+455923.4 (hereafter J0821), which is
an eclipsing binary with an orbital period of $P_{orb}\sim 0.51$~day,  a  WD's surface temperature of $T_{eff}\sim 80000$~K and M-type companion star \citep{2013MNRAS.429..256P}. Since these  binary parameters  are similar to those of ASJ2005, the detection of the short periodic signal is expected. However,
the data taken with 120-second cadence model of TESS does not show a
significant periodic signal shorter than the orbital period.
In spite of edge-on view, the ZTF light curve shows the amplitude of the orbital modulation is $\sim 0.7$ magnitudes, which is smaller than $\sim 1.5$ magnitudes of ASJ2005.  This indicates that the optical emission from J0821 is more dominated by the WD emission
than that from ASJ2005. A deeper observation for J0821 is desired to search for  the short periodic
modulation and to study the  similarity/dissimilarity with   ASJ2055.

In summary, we carried out the photometric study for ASJ2055, whose optical emission shows
(i) an orbital modulation ($P_{orb}\sim 0.52$~hours)
with an amplitude of $\triangle m\sim 1.5$~magnitude and (ii) a $9.77$-minute modulation. With TESS data, we found that
(i)  the period of the short modulation  measured in 2022 August  is significantly smaller ($|\triangle P_0|/P_0\sim 8\times 10^{-4}$) than that measured in 2021 July/August
and (ii) TESS 2022 August data also evolution of the periodic signal with
the a time derivative of $\dot{P}_0=1.2(5)\times 10^{-7}$.
This large variability of the  9.77-minute periodic signal 
will be    incompatible with the scenario  of the  WD's spin. Alternatively,  the oscillation of M-type star is more likely as the origin of the periodic modulation.
The optical/UV spectrum and orbital modulation suggest that ASJ2055 is a PCEB and the radiation from a hot WD heats up the day-side of the companion star.  ASJ2055 may be a new type of the WD binary system, in which a hot WD heats up a oscillating M-type star.

\vspace{4mm}
We thank to referee for his/her useful comments and suggestions.
We are grateful to  Swift-TOO team for arrangements of observations for our sources. J.M. thanks the discussion with
Prof. Yan Li and Xiangdong Li. J.T. and X.F.W are supported by 
the National Key Research and Development Program of China (grant No. 2020YFC2201400) and the National Natural Science Foundation of China (grant No. 12173014). A.K.H.K. is supported by the National Science and Technology Council of Taiwan through grant 111-2112-M-007-020.  J.M. is supported by the National
Natural Science Foundation of China (grant No. 11673062). X.H. is supported by the National Natural
Science Foundation of China through grant 12041303 and Yunnan Revitalization Talent Support Program (YunLing Scholar Award).
 C.-P.H. acknowledges support from the National Science and Technology Council  through MOST 109-2112-M-018-009-MY3. L.C.-C.L. is supported by  NSTC through grants  110-2112-M-006-006-MY3 and 111-2811-M-006-012. K.L.L. is supported by the National Science and Technology Council  through grant 111-2636-M-006-024, and he is also a Yushan Young Fellow supported by the Ministry of Education of the Republic of China (Taiwan).  C.Y.H. is supported by the research fund of Chungnam National University and by the National Research Foundation of Korea grant 2022R1F1A1073952

 \bigskip
 
 \section*{Data  Availability}
 All data and analysis products presented in this article are available upon request. The TESS data used in this paper can be found in MAST \citep{TESSALL}. The GALEX, Pan-STARRS and WISEdata used in this papercan be found inVizieR (\url{https://vizier.cds.unistra.fr/viz-bin/\
   VizieR}). The Siwft data used in this paper can be obtain from NASA's HEASARC Archive (\url{https://heasarc.gsfc.nasa.gov/docs/archive.html}).

\facility{{\it Swift}(XRT), {\it ZTF}, {\it TESS}~{\it LT}  and {\it LOT}}.

\software{
\newline {\tt Ximage} \\\url{https://heasarc.gsfc.nasa.gov/xanadu/ximage/ximage.html}
\newline {\tt IRAF} \\ \url{https://iraf-community.github.io}, \citealt{1993ASPC...52..173T}
 }

\bibliography{ref}

\end{document}